\documentclass[sigconf]{acmart}

\AtBeginDocument{%
  }

\setcopyright{acmlicensed}
\copyrightyear{2024}
\acmYear{2024}
\acmDOI{XXXXXXX.XXXXXXX}

\acmConference[SA Art Papers '24]{SIGGRAPH Asia 2024 Art Papers}{December 03--06, 2024}{Tokyo, Japan}

\acmBooktitle{SIGGRAPH Asia 2024 Art Papers (SA Art Papers '24),
 December 03--06, 2024, Tokyo, Japan}
\acmISBN{978-1-4503-XXXX-X/18/06}

\begin{document}

\title{Exploring Fungal Morphology Simulation and Dynamic Light Containment from a Graphics Generation Perspective}


 \author{Kexin Wang}
 \email{kexin_wang3@brown.edu}
 \orcid{0009-0001-2835-4148}
 \affiliation{%
   \institution{Brown University, School of Engineering
   \\Rhode Island School of Design}
   \city{Providence}
   \state{Rhode Island}
   \country{USA}
 }

\author{Ivy He}
\email{xiao_he@brown.edu}
\affiliation{%
  \institution{Brown University, Department of Computer Science}
  \city{Providence}
  \state{Rhode Island}
  \country{USA}
}

\author{Jinke Li}
\email{Jinke_li@brown.edu}
\affiliation{%
  \institution{Brown University, Department of Molecular Microbiology and Immunology}
  \city{Providence}
  \state{Rhode Island}
  \country{USA}
}

\author{Ali Asadipour}
\email{ali.asadipour@rca.ac.uk}
\orcid{0000-0003-0159-3090}
\affiliation{%
  \institution{Royal College of Art, Computer Science Research Centre}
  \city{London}
  \country{United Kingdom}
}

\author{Yitong Sun}
\email{yitong.sun@network.rca.ac.uk}
\orcid{0000-0002-9469-7157}
\authornote{Corresponding author.}
\affiliation{%
  \institution{Royal College of Art, Computer Science Research Centre}
  \city{London}
  \country{United Kingdom}
}

\renewcommand{\shortauthors}{Kexin Wang et al.}

\begin{abstract}
Fungal simulation and control are considered crucial techniques in Bio-Art creation. However, coding algorithms for reliable fungal simulations have posed significant challenges for artists. This study equates fungal morphology simulation to a two-dimensional graphic time-series generation problem. We propose a zero-coding, neural network-driven cellular automaton. Fungal spread patterns are learned through an image segmentation model and a time-series prediction model, which then supervise the training of neural network cells, enabling them to replicate real-world spreading behaviors. We further implemented dynamic containment of fungal boundaries with lasers. Synchronized with the automaton, the fungus successfully spreads into pre-designed complex shapes in reality.
\end{abstract}


\begin{teaserfigure}
  \includegraphics[width=\textwidth]{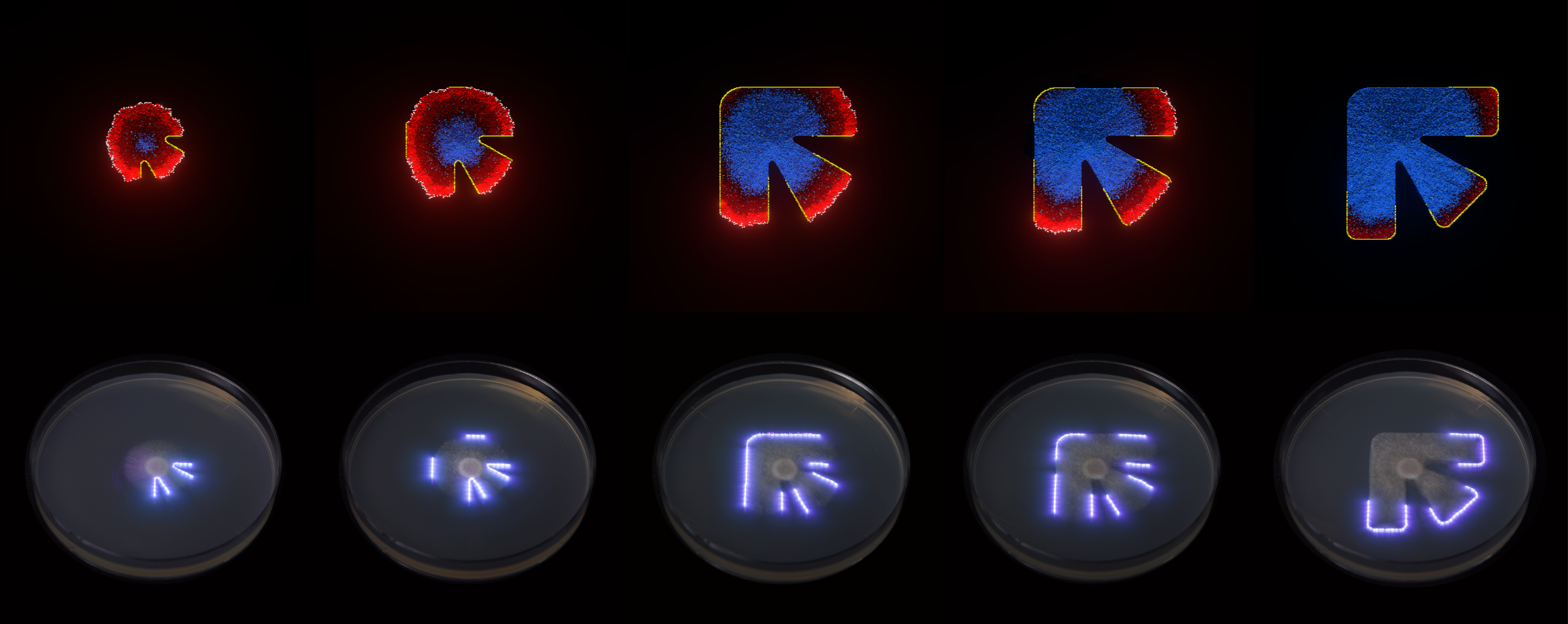}
  \caption{Simulation of fungal spread using our neural network-based fungal automaton, supervised and trained by an AI model chain. It is then synchronized with a laser in reality to implement dynamic light containment, guiding fungal growth to match pre-designed patterns consistent with the simulation.}
  \Description{Teaser figure}
  \label{fig:teaser}
\end{teaserfigure}


\maketitle

\section{Introduction}
The simulation and control of fungal morphology has garnered increasing interest in fields such as art, gaming, architecture, design, and human-computer interaction (HCI) \cite{kim2023surfacing}. In recent years, deeper integration of art with biological research has gradually clarified the concept of Bio-Art. As fungal research becomes more entrenched in Bio-Art, it continues to spark new inquiries into ethics, social issues, and aesthetics, while also providing unique inspiration for other fields such as robotics, sensor research, and bio-manufacturing \cite{wicaksono2023fungi}.

Although practitioners of fungal art are often classified as artists, creating art based on fungal morphology typically requires a deep theoretical understanding of biology. The integration of computer algorithms and simulation technologies has bridged the gap between artists and fungal morphology, significantly reducing the time consumption and validation difficulties of traditional fungal experiments. However, for most artists, operating computer programs remains a significant challenge.

Our research treats the fungal spread pattern as a two-dimensional graphic time-series generation problem. We propose a simulation pipeline based on a neural network-driven fungal cellular automaton. This pipeline first employs the Efficient-ViT (E-ViT) image segmentation model and a customized Temporal Convolutional Network (TCN) time-series model to learn the patterns of fungal spread from sequences of images \cite{cai2022efficientvit, lea2017temporal}. The TCN model then provides supervised training to the neural network (NN) cells, enabling each cell to independently respond in a manner consistent with the learned fungal spread patterns, thereby achieving realistic morphology simulations. Furthermore, to achieve pre-designed spread shapes, we experimentally introduce the use of lasers to establish dynamic light boundaries that limit the direction of fungal growth. The automaton is then connected to a real-world laser device, which successfully guides the fungus to spread into various complex shapes during practical tests, in accordance with the simulation predictions, without the need for computer vision (CV) monitoring.

Benefiting from the pipeline of AI models and the design of the NN-based cellular automaton, artists can simply provide a video of fungal spread to achieve high-fidelity simulations of fungal morphology and real-world light control, without needing to delve into programming and algorithms.

\section{Related Works}
\subsection{Fungal Morphology in Art}
Artists and researchers have created numerous works through the study of fungal morphology. Groutars et al. explored using Flavobacteria in HCI interfaces, focusing on their controllable color display during spreading changes \cite{groutars2022flavorium}. Kim et al. developed a hybrid bio-digital game that investigates the ethical impacts and significance of fault events by observing interactions among humans, computers, and mold \cite{kim2019moldy}. Gianluca et al. applied parametric modeling techniques to study the aggregation effects and growth strategies of fungi, exploring their potential in bio-architectural structures \cite{Tabellini2015}.

Although these works utilized various technologies to simulate fungal spreading, they still required complex computer programming and algorithm development. In contrast, our approach employs AI models to learn fungal growth patterns from image sequences and supervises the training of an NN-based cellular automaton. This method allows users to conduct morphology studies with zero coding, significantly reducing the reliance on complex programming and algorithms.

\subsection{Fungal Cellular Automata}
In fungal biology research, cellular automata are extensively employed as an efficient tool for studying and simulating fungal growth and spread behaviors. Andrew et al. developed basic spreading rules for cellular automata from microscopic studies of Woronin bodies in fungal hyphae \cite{adamatzky2023fungal}. Iuri et al. reduced fungal growth rules to four states to facilitate simulations of various scenarios \cite{de2013modelling}. Runyu et al. explored how transitions from simple to complex rules impact the performance of fungal cellular automata \cite{zhang2022multi}. Additionally, artworks by Cesar and Lois, showcased at Siggraph, used simple-rule-based cellular automata to depict the erosion of a physical book by virtual and real fungi \cite{CesarLois2018}.

These studies underscore the versatility and utility of cellular automata in fungal simulations, typically starting from a microscopic perspective and iterating based on static rules. However, such methods have limitations, especially in capturing the complexity and dynamic changes in the system as the fungal mass increases. By contrast, our NN-based cellular automata dynamically adjust parameters during iterations, learning spreading rules from macroscopic real-world data. This method allows users to bypass the need for intricate microscopic understanding, with spreading patterns automatically inferred by the AI model pipeline.

\subsection{Fungal Spread Control}
Fungal spread can be influenced by various external factors. Yu et al. explored the impact of food baits on the spread of Phanerochaete velutina, demonstrating how attractants can alter fungal behavior \cite{fukasawa2020ecological}. Matilde et al. assessed the effects of magnetic fields on the spreading behaviors of three different fungi, revealing how non-chemical stimuli influence growth patterns \cite{anaya2021effect}. Chien-Wei et al. utilized multicolored LED lights to regulate the growth of Aspergillus ficuum, indicating the potential of light spectra in fungal cultivation \cite{cheng2012effect}. Graham et al. developed an "optical tweezers" technique to precisely guide the movement of hyphae at the microscopic level \cite{wright2007optical}.

In our research, we initially evaluated the containment effects of lasers with varying wavelengths and power levels on fungal spread. We then integrated this containment technique with our simulation system, establishing dynamic light boundaries that guide the fungus into pre-designed complex shapes. Additionally, our simulation engine's modular design enhances its adaptability, allowing for future integration with various inducing factors.

\begin{figure}[h]
  \centering
  \includegraphics[width=1.0\linewidth]{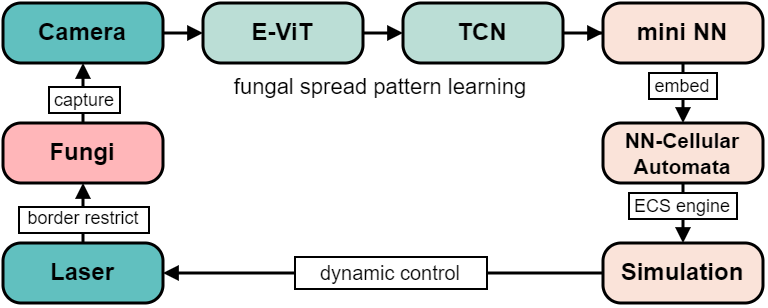}
  \caption{Framework of our method. The E-ViT model first captures temporal edge segmentation images by observing real fungal spread. Subsequently, the TCN model learns spread patterns from these temporal images. Third, the TCN model is used for supervised training of the NN that drive the cellular automaton, aligning it with real-world fungal growth patterns. Finally, the laser device used to limit the spread is connected to the NN-cellular automaton, facilitating the execution of simulations while achieving the spread of pre-designed complex shapes.}
  \label{fig:framework}
\end{figure}

\section{Methodology}
Our study primarily explores two approaches for the controlled simulation of fungal morphology: NN-cellular automata simulation and light containment. The framework of our methods is depicted in Figure \ref{fig:framework}.

\subsection{Concept of Simulation}
Fungal spread and growth can be viewed as a complex system. Previous research has employed various methods to model and simulate this process, with cellular automata achieving satisfactory results at micro scales. However, as the scale of growth expands, the complexity increases dramatically, rendering cells with fixed parameters inadequate for capturing changes in fungal spread patterns.

Our approach conceptualizes the macroscopic spread of fungi as a 2D graphic time-series generation problem. Specifically, we utilize a chain of machine learning models to learn parameters from real sequences of fungal spread images. These parameters are then integrated into a compact NN, which dynamically adjusts the behavior patterns of each cell based on varying conditions within the simulation. This method allows us to accurately simulate fungal morphology at a macro scale, closely approximating real-world patterns.

Benefiting from this innovative pipeline design, our approach enables artists and interdisciplinary researchers to participate in fungal spread simulations without the need for extensive mathematical knowledge or coding skills. By merely providing a sequence of fungal growth images, our system uses NN-driven cells that autonomously adapt, seamlessly integrating the simulation into their projects.

\subsection{Learning from Fungal Image Sequences}
To study and replicate the patterns of fungal spread, we developed a comprehensive pipeline starting from an image segmentation model to a time-series prediction model.

\begin{figure}[t!]
  \centering
  \includegraphics[width=1.0\linewidth]{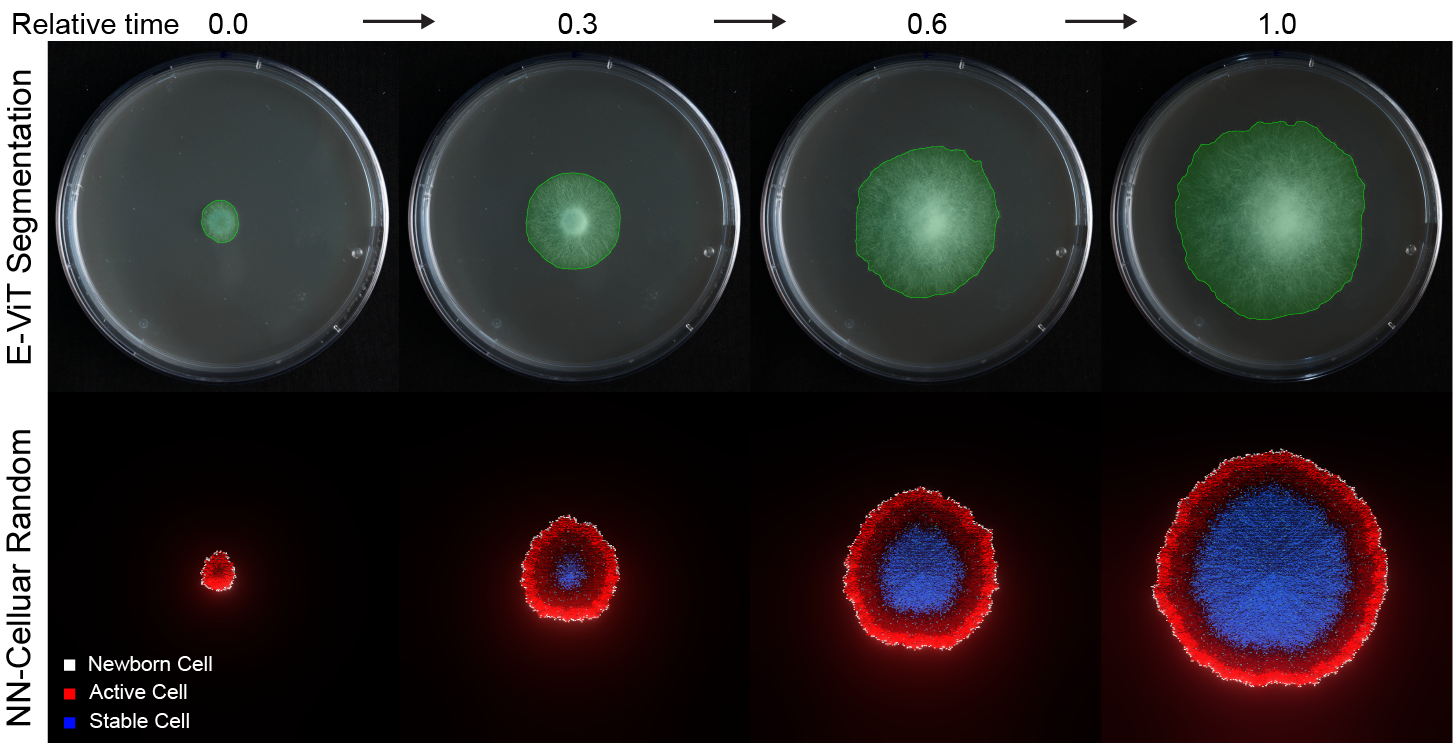}
  \caption{AI model chain learn fungal spread pattern and supervise the training of NN-cell for simulation. Upper: E-ViT captures the edge of sequential fungal images and transfers them to the TCN model to learn its spreading rules. Lower: The TCN model supervises the training of NN-cells, enabling them to execute fungal spread patterns in the simulation environment that are consistent with reality.}
  \label{fig:evit_learning}
\end{figure}

Initially, we selected the pre-trained E-ViT-SAM-L0 image segmentation model as the starting point of our pipeline. This model has been then fine-tuned particularly for handling the blurring issues at the edges of fungi. As shown in Figure \ref{fig:evit_learning} upper, it accurately identifies fungal contours under diverse lighting and color conditions, even when the contrast between the fungus and the background is low. Additionally, E-ViT is designed to operate efficiently on low-power devices, significantly reducing the implementation barriers of our pipeline.

Subsequently, we designed a specialized TCN model with three convolutional layers to predict fungal growth patterns from the contour changes segmented by the E-ViT model. The network processes five sequential images of 512x512 resolution through a sliding window temporal model, outputting a predicted spread image of the same resolution. It starts with a 2D convolutional layer using a (3x3) kernel size and (2,2) stride for down-sampling. This is followed by a series of 1D temporal convolutional layers that handle time-dependent relationships, employing a kernel size of 3, a stride of 1, and progressively increasing dilation factors (1, 2, 4) to enhance the receptive field and capture information over longer time spans. Batch normalization follows each convolutional layer to stabilize the training process and improve generalization. The ReLU activation function is used to increase the network’s non-linear processing capabilities. The output layer comprises three transposed convolutional layers that incrementally up-sample the feature map back to the original resolution. The model is trained using the Adam optimizer with mean squared error as the loss function, incorporating dropout where necessary to mitigate over-fitting. Once trained, the model can predict the future contour of fungal spread based on data from the past five moments, visually representing the rules of fungal spreading from an image generation perspective. This model will subsequently be used to train and supervise the NN-cellular automata.

\begin{figure*}[t!]
 \centering
 \includegraphics[width=\textwidth]{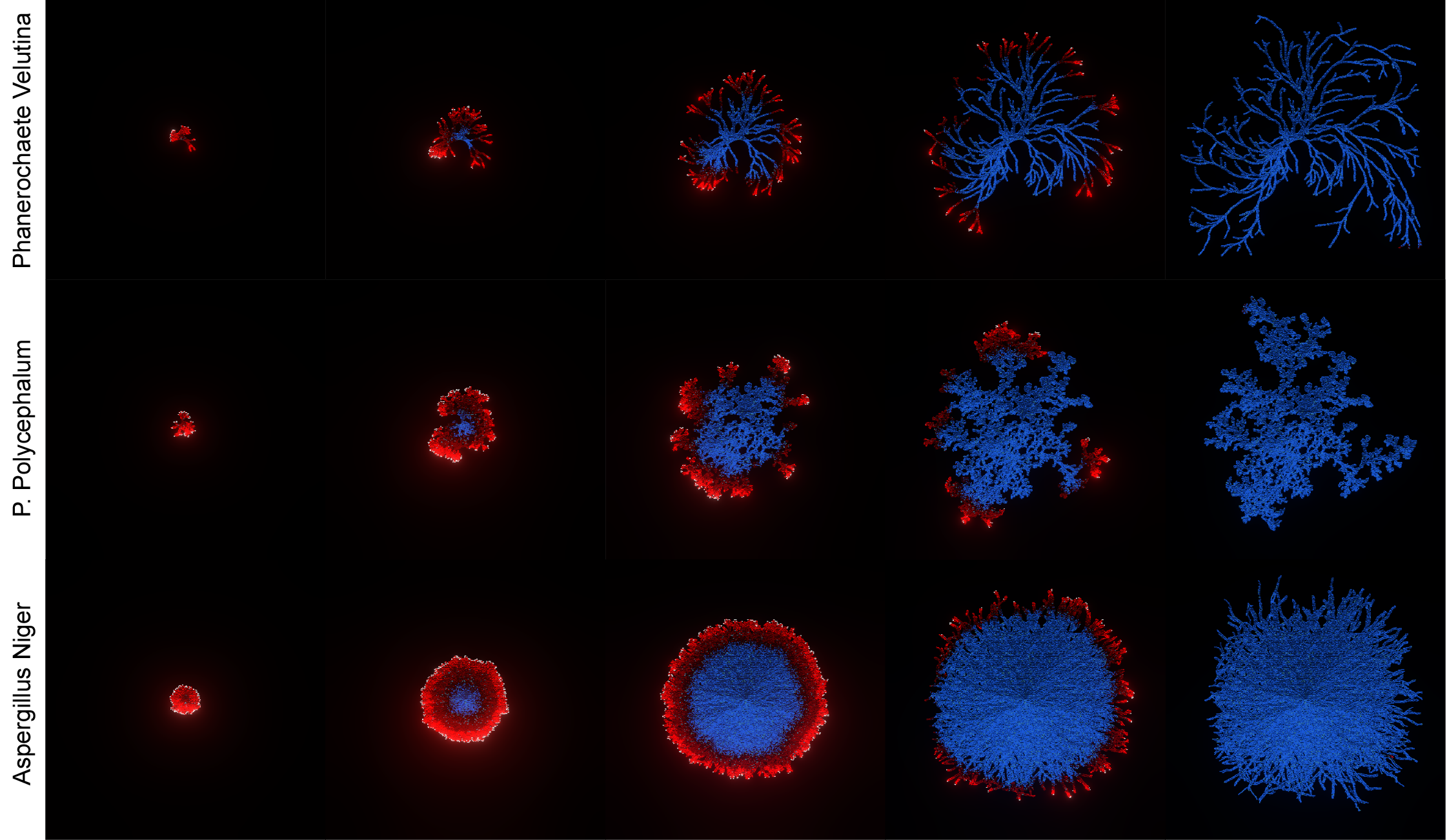}
 \caption{Random spreading simulations of Phanerochaete velutina, Physarum polycephalum, and Aspergillus niger.}
 \label{fig:three_fungi}
\end{figure*}

\subsection{Neural Network Fungal Cellular Automata}
Traditional cellular automata typically utilize a set of simple rules to generate complex emergent behaviors. However, in simulations with rapidly increasing cell numbers, these static rules often fail to capture the complexities of real fungal spread. To address this, our approach integrates a NN within each cell, replacing limited rule parameters with more flexible, dynamic ones to enable complex fungal behaviors in large-scale simulations. For efficient computation, we developed a low-latency 2D game engine based on the Entity-Component-System (ECS) architecture in Rust, enhanced with Vulkan for GPU acceleration and supporting multi-threading on CPUs, facilitating efficient cross-platform real-time fungal simulation \cite{harkonen2019advantages}. The standard ECS architecture also simplifies future modifications for game developers.

In our cellular architecture, each cell "entity" includes five "components": viability state, nutrient residue, division count, generation number, and coordinates. During each iteration, a neural network system queries all existing cell "entities" within the simulation environment, setting up a 3x3 receptive field to sense the immediate neighbors' states and subsequently update the current cell’s state after NN processing. The feed-forward NN's input layer comprises 45 features (5 features each from the cell and its 8 neighbors), two hidden layers with 90 and 60 neurons respectively, using ReLU activation. The output layer, activated by sigmoid functions, has 5 neurons. Supervised training of this NN is performed using fungal image predictions from the TCN model as labels, representing the "ideal" future state for each cell. By minimizing cross-entropy loss between the outputs of the cellular automaton and the TCN predictions, the neural network effectively simulates the TCN's global predictions based on local information. This approach overcomes the TCN's limitations of fixed simulation size and dataset dependency, allowing for the introduction of additional programmatic rules and external constraints such as boundary conditions, food source guidance, and interactions among multiple fungal species, thereby enriching the simulation’s detail. Figure \ref{fig:evit_learning} demonstrates an example of this process, showing the progression from edge detection by E-ViT to random simulations in our neural network-driven environment after learning the spreading patterns of Rhizopus Oligosporus.

Performance tests were conducted on a 2021 Razer Blade laptop with an RTX3070 GPU. The simulation environment maintained a refresh rate of 180Hz with up to one hundred thousand cells, and 30Hz with one million cells, showcasing the computational benefits of our Rust-based ECS engine and enabling consumer-grade computers to perform large-scale real-time simulations of fungal morphology.

\subsection{Verification Across Fungal Species}
To verify the effectiveness of our method across different fungal species, we selected three fungi with distinct spreading patterns for simulation: Phanerochaete velutina, known for its linear branching structure; Physarum polycephalum, characterized by its interwoven network structure; and Aspergillus niger, noted for its uniform spread \cite{tlalka2008quantifying, latty2009food, baker2006aspergillus}. We sourced sequence images of these fungi from multiple open-source online databases, using them as raw data for our pipeline. The NN-cells were then trained with specific parameters derived from this data. Figure \ref{fig:three_fungi} showcases the results of random spreading simulations for each of the three fungi. The results confirm that our method can not only effectively learn and replicate the natural spreading patterns of each species but also capture the inherent randomness of complex systems. Importantly, the spreading patterns remained stable despite a rapid increase in cell numbers.

\begin{figure}[t!]
  \centering
  \includegraphics[width=1.0\linewidth]{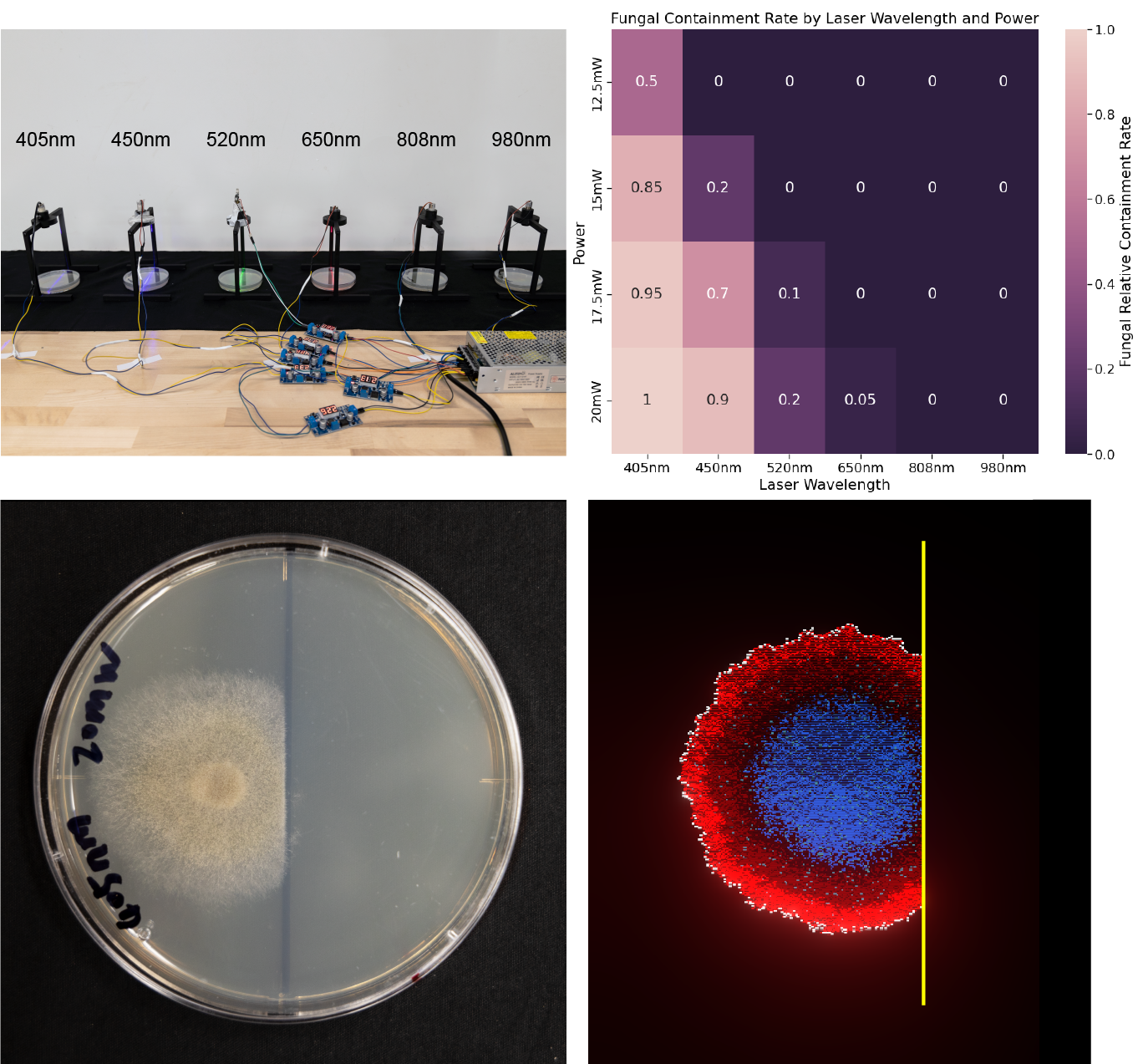}
  \caption{Experiment on light containment of Rhizopus Oligosporus. Upper left: Experimental setup. Upper right: Normalized containment rates of Rhizopus Oligosporus with lasers of different powers and wavelengths. Lower left: Detail of containment using a 405 nm 20 mW laser. Lower right: Simulated light containment using NN-cellular automata.}
  \label{fig:compare}
\end{figure}

\begin{figure}[t!]
  \centering
  \includegraphics[width=1.0\linewidth]{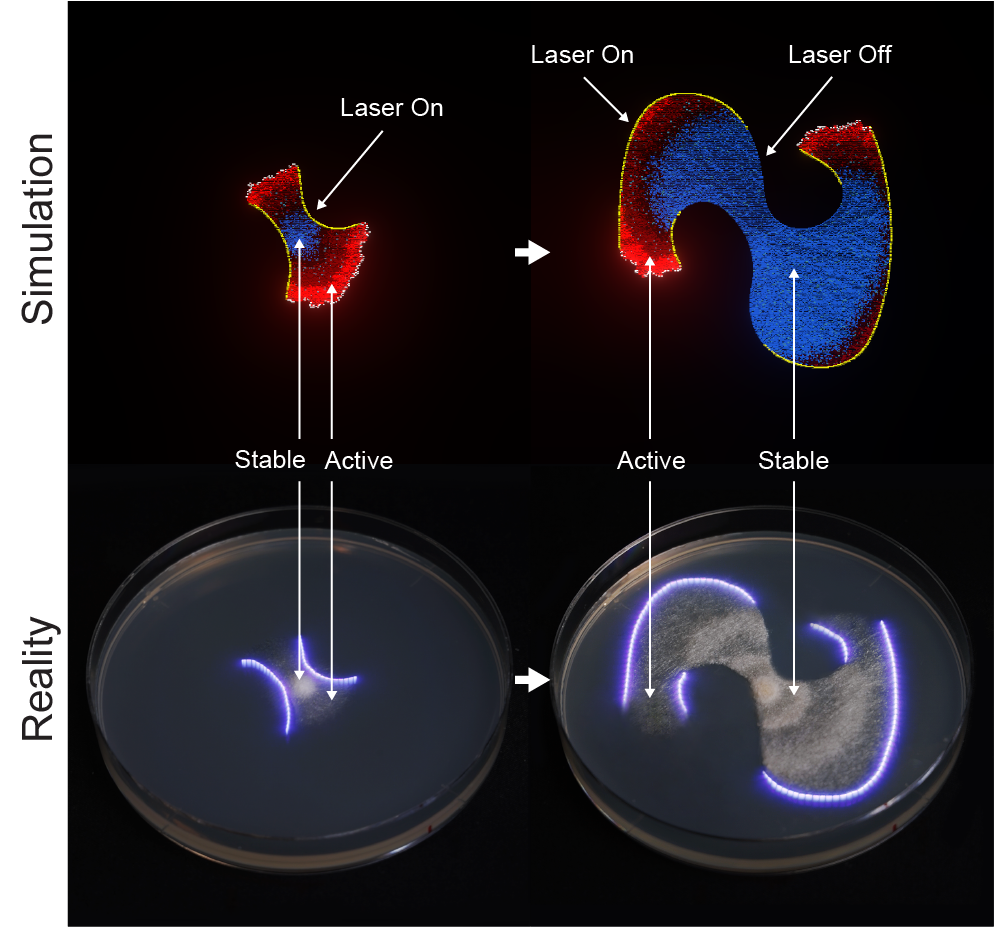}
  \caption{Dynamic light containment testing. Upper: The NN-cellular automaton, with the addition of light entities, achieves dynamic light containment, turning off in stable fungal areas and on in active areas. Lower: The laser device, connected to the simulation environment via an API, successfully forms fungal spread patterns consistent with the simulation after executing a dynamic light containment strategy that mirrors the simulation.}
  \label{fig:dyn_light}
\end{figure}

\subsection{Containment of Fungi by Light}
While we have successfully conducted free spreading simulations of fungi, achieving the precise control needed for artistic and research applications—shaping fungi into specific forms—remains challenging. In the related work section, we reviewed various controlling methods. We propose a method that combines the establishment of linear light boundaries with NN-driven cellular automata to achieve controllable fungal behavior both in virtual and real environments.

In practical tests, we used lasers of specific wavelengths to create light boundaries that effectively contained the fungal growth. To elucidate the relationship between laser wavelength, light energy, and fungal behavior, we selected six readily available laser types, ranging from blue visible light to infrared (405 nm, 450 nm, 520 nm, 650 nm, 808 nm, and 980 nm), to assess their containment efficacy on Rhizopus Oligosporus. We prepared fungal spores in a 1:3 weight ratio with water, with 15 micro-liters used for each inoculation on Potato Dextrose Agar (PDA) within a constant 24 degrees Celsius environment. Lasers were calibrated using an ThorLab S320C power meter at four power settings, and with equal luminous width: 12.5 mW, 15 mW, 17.5 mW, and 20 mW. Experiments were conducted over 48 hours and repeated five times to average individual variations in fungal spread. After each experiment, hyphal counts per unit area under illumination were recorded using a microscope to calculate the containment rate. The containment rate is inversely proportional to the hyphal counts per unit area.

The normalized experimental data are illustrated in the upper right of Figure \ref{fig:compare}, show that shorter wavelengths generally provided higher containment effectiveness at the same energy levels, with wavelengths over 520 nm proving almost ineffective. Notably, a 405 nm laser at 20 mW achieved complete containment, while other settings exhibited some leakage. Thus, a 20 mW 405 nm laser was determined to be the most effective for controlling Rhizopus Oligosporus growth.

After determining the optimal light boundary parameters, we integrated decision logic into the neural network cycle within the simulation environment to mimic the light boundary containment effects. Figure \ref{fig:compare} compares the real petri dish results using a 405 nm laser at 20 mW with virtual environment simulations, demonstrating the method's effectiveness.

\begin{figure*}[t!]
 \centering
 \includegraphics[width=\textwidth]{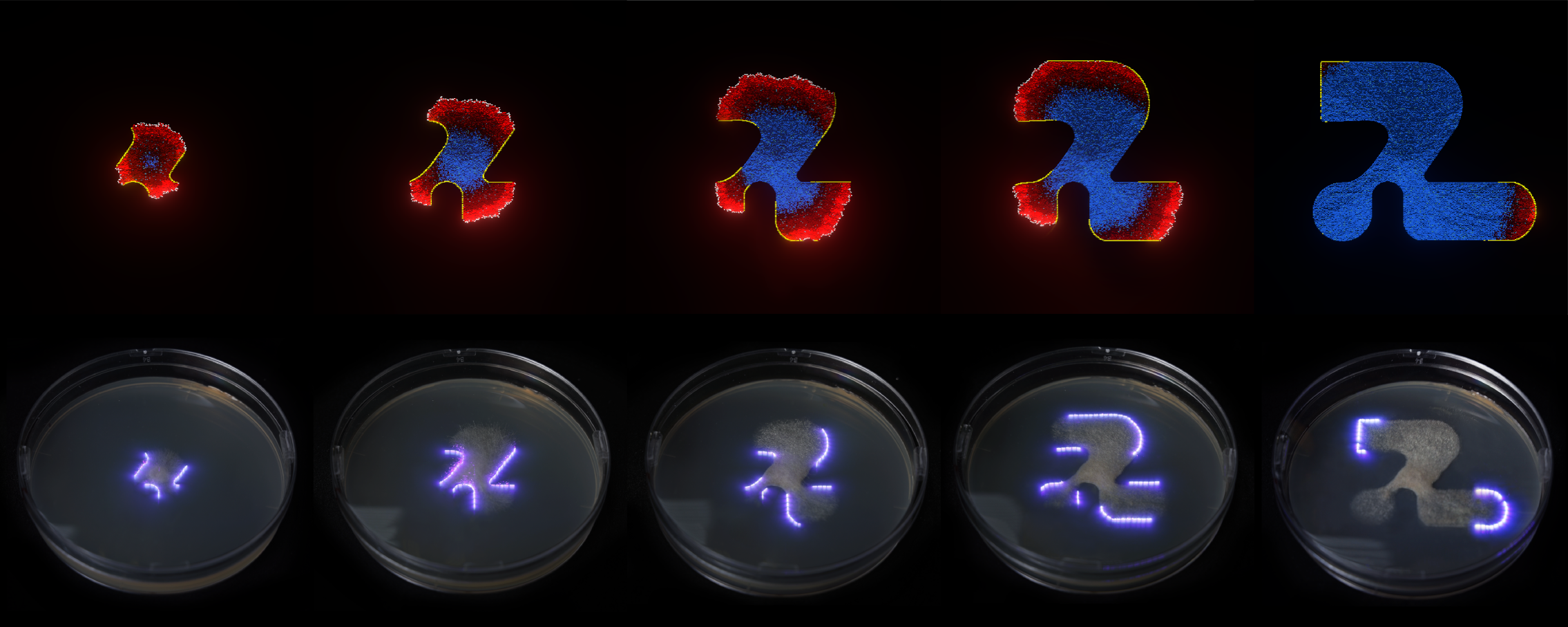}
 \caption{Comparison of simulated and actual fungal spread under synchronized dynamic light control. Upper: Fungal spread and dynamic light boundary changes simulated by the NN-cellular automaton. Lower: Actual fungal spread contained by laser in a real-world setting.}
 \label{fig:impl}
\end{figure*}

\subsection{Dynamic Light Control in Simulation}
Following the light boundary experiments on Rhizopus Oligosporus, we noted that fungal regions depleting their nutrients entered a stable state, halting their spread. Conversely, regions with ample nutrients remained in an "active state," continuing to spread. These observations led us to hypothesize whether activating lasers only at the edges of the active regions and turning them off at the stable regions could achieve the same containment effect as activating lasers along the full boundary. This strategy could potentially allow for unlimited canvas control over fungal spread and enhance energy efficiency in applications.

To test the feasibility of dynamic light containment, we refined our simulation engine. Initially, we introduced light "Entity" cells, analogous to NN fungal cells, but designed to respond based on their coordinates. After each iteration, these light cells automatically distinguish and categorize active from stable fungal areas, adjusting their "on" or "off" states accordingly. We then applied the Breadth-First Search (BFS) algorithm to organize the "on" state light cells into required line segment combinations. Subsequently, we converted the laser lines into vector data and developed an application programming interface (API) for direct G-code transmission to the laser device, replicating the established laser boundary in the actual environment.

Figure \ref{fig:dyn_light} illustrates an example of dynamic light containment directing controlled fungal spread. The real-world experiments confirmed our hypothesis's validity. Without relying on machine vision for growth monitoring, our neural network cellular automaton precisely predicted fungal spread changes. By synchronously controlling the laser with the G-code API and focusing the boundaries solely on active fungal areas, we effectively managed the spread to mimic complex patterns as predicted by the simulation.

\section{Implementation}
\subsection{Minimum Power Consumption Fungal Spreading Control}
To illustrate the efficacy of our method, we designed an example focused on controlling fungal simulations with minimal light power consumption. This example serves as an inspiration for artists, designers, and researchers interested in low-power projects such as sustainable fungal fabrics, food design, construction of structures in outer space, and the production of fungal flexible circuit boards.

We used Rhizopus Oligosporus in our experiments due to its broad applications in food fermentation, waste treatment, and as an industrial enzyme for breaking down fats and proteins\cite{Duygu2008Use}. The shapes designed for testing—curves, narrow passages, arcs, and straight lines—were chosen to comprehensively represent the diverse challenges encountered by fungal spread in various environments. To minimize power consumption, an annealing algorithm was employed to optimize fungal inoculation points through simulation iterations before actual spreading. Once the optimal starting points were established, we synchronized the fungal spread rate in the simulation environment with real-world conditions.

The laser system, integrated via an API, was controlled in real-time using G-code. Figure 8 presents a comparison between simulated and actual fungal spread, emphasizing the adaptation of laser boundaries over time with dynamic light containment strategies. Test results confirmed that our NN-based cellular automaton successfully learned and replicated the spreading patterns of Rhizopus Oligosporus, accurately mirroring the predetermined shapes under dynamic light control strategies.

\begin{figure*}[t!]
 \centering
 \includegraphics[width=\textwidth]{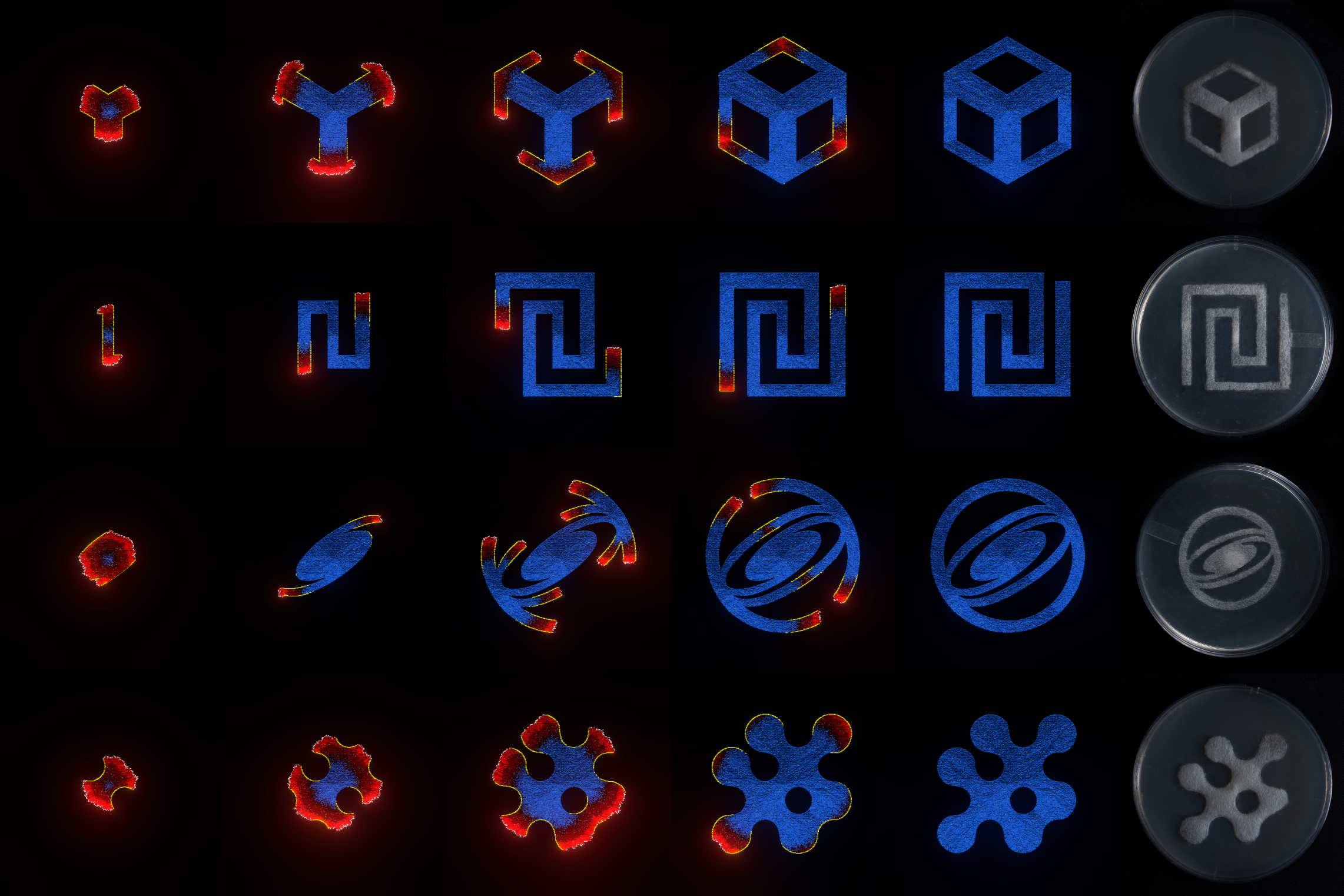}
 \caption{NN fungal cellular automaton simulations of four complex shapes with dynamic light containment in reality.}
 \label{fig:four_pattern}
\end{figure*}

\subsection{Multi-contour Spreading Test}
Building on our previous successes, we challenged our method with more complex shapes to assess its effectiveness on targets featuring finer details, extended growth periods, and intricate structures. We selected four distinct shapes for testing: a hollowed-out form, a looping corridor, a 2D Logo, and a network structure. Figure \ref{fig:four_pattern} shows the simulation process of these shapes using the NN-cellular automaton, including dynamic light control details, and showcases the final fungal formations in actual petri dishes. The results confirm that our method accurately replicates the intended boundaries of these complex shapes, effectively managing the fungal growth without significant overgrowth, demonstrating the precision of our fungal spread predictions and dynamic laser control.

\section{Conclusion}
This study demonstrates the powerful capabilities of combining NN-based cellular automata with dynamic light control strategies to accurately simulate and control fungal morphology. By establishing a AI model chain, our approach has successfully automated the learning of fungal spread complexities, not only replicating natural patterns with high precision but also precisely shaping fungal growth into complex geometries as required by real-world applications. By integrating the complex pipeline into an ECS game engine, we have provided a robust zero-coding simulation tool for artists, designers, and researchers looking to explore and utilize fungal growth in sustainable materials, bio-art, and other cutting-edge applications. Moreover, our engine is designed modularly, allowing developers to easily reassemble it to enhance NN-cells for specific uses. We plan to open-source our project code to promote the development and application of fungal morphology simulation technologies in related fields.

\section{Future Work}
For next step, we plan to enhance the usability of our method in three key areas: First, we will develop a user-friendly UI that not only provides more intuitive parameter adjustments but also increases ease of use. Second, we will optimize our algorithms and collect and construct a database of fungal temporal spread to support broader applications and research. Finally, we intend to extend our method, which combines the AI pipeline and NN-cell automata, to other areas, for instance, fire spread simulation, to explore its potential in other scenarios.

\bibliographystyle{ACM-Reference-Format}
\bibliography{reference}


\end{document}